\documentclass[aps,prx,twocolumn,amsmath,amssymb,10pt]{revtex4-2}
\usepackage{ragged2e}
\usepackage{blindtext}
\usepackage{graphicx}
\pdfoutput=1
\usepackage{dcolumn}
\usepackage{bm}
\usepackage{multirow}
\usepackage{float}
\usepackage{wrapfig}
\usepackage[normalem]{ulem}
\usepackage{color}
\usepackage{xr}
\usepackage{booktabs}
\usepackage{longtable}
\usepackage[colorlinks=true, citecolor=blue, urlcolor=black]{hyperref}
\usepackage{xcolor,calc}
\hypersetup{linkcolor = black, citecolor = black, urlcolor = blue} 
\usepackage{mdframed}
\usepackage{graphicx}
\usepackage{dcolumn}
\usepackage{bm}
\usepackage{fancybox}
\usepackage{amsmath}
\usepackage{enumerate}
\usepackage{pslatex}
\usepackage{relsize}
\usepackage{setspace}
\usepackage[utf8]{inputenc}

\usepackage[T1]{fontenc}
\usepackage{lmodern}
\usepackage{hyperref}
\usepackage{verbatim}
\usepackage{amssymb}
\usepackage[toc,page]{appendix}
\usepackage[version=3]{mhchem}
\usepackage{chemfig}
\usepackage{numprint}
\usepackage{calc}
\usepackage{fancyhdr}
\usepackage{array}
\usepackage{xcolor}
\newcolumntype{P}[1]{>{\centering\arraybackslash}p{#1}}

\usepackage{soul} 

\DeclareUnicodeCharacter{2010}{-}

\begin{document}

\title{Quantitative modeling of spintronic terahertz emission due to ultrafast 
spin transport}
\author{Francesco Foggetti}
\author{Peter M. Oppeneer}
\affiliation{Department of Physics and Astronomy, P.\,O.\ Box 516, Uppsala University, SE-751 20 Uppsala, Sweden}

\date{\today}

\begin{abstract}
In spintronic terahertz emitters, THz radiation is generated by exciting an ultrafast spin current through femtosecond laser excitation of a ferromagnetic-nonmagnetic metallic heterostructure. 
Although an extensive phenomenological knowledge has been built up during the last decade, a solid theoretical modeling that connects the generated THz signal to the laser induced-spin current is still incomplete. 
Here, starting from general solutions to Maxwell's equations, we model the electric field generated by a superdiffusive spin current in spintronic emitters, taking Co/Pt as a typical example. We explicitly include
the detector shape  which is shown to significantly influence the detected THz radiation. Additionally, the electron energy dependence of the spin Hall effect is taken into account, as well as the duration of the exciting laser pulse and thickness of the detector crystal. 
Our modeling leads to realistic emission profiles and highlights the role of the detection method for distinguishing key features of the spintronic THz emission.
\end{abstract}

\maketitle

\section{Introduction}

The terahertz (THz) region is defined as the interval in the electromagnetic spectrum with frequencies in  the range of about 0.3 to 30 THz \cite{Ferguson2002,Tonouchi2007,Lee2009,Dexheimer2017}.  As the number of possible THz-based applications continues to grow, spanning diverse fields such as imaging, security, communication, and high speed electronics \cite{Tonouchi2007,Mittleman2017,Chen2020}, the last decade has been characterized by a great effort to realize efficient THz emitters \cite{Seifert2016,Dhillon2017,Papaioannou2018,Chen2019,Papaioannou2020}.  

Current state-of-the-art THz emitters are nonlinear semiconductor crystals that emit THz radiation due to optical rectification or, alternatively,  photoconductive antennas \cite{Nahata1996,Burford2017,Dhillon2017,Lee2009}. However, the limited bandwidth of these emitters has motivated  investigations in alternative solid-state emitters. A few years ago, spintronic THz emitters were discovered as a promising class of emitters that could offer a wide bandwidth of ten or more THz \cite{Kampfrath2013,Seifert2016,Wu2017,Adam2019,Dang2020,Papaioannou2020,Wu_principles_2021,Bull2021,Feng2021,Kolejak2024}.

The general structure of a spintronic THz emitter consists of a bilayer of a ferromagnetic (FM) metal and a nonmagnetic (NM) heavy metal, with layer thicknesses of several nanometers \cite{Seifert2016,Wu2017,Adam2019,Dang2020,Papaioannou2020,Wu_principles_2021,Bull2021,Feng2021}.   Despite this simple structure, and although significant experimental studies have been performed, the physical origin of the THz emission continues to be disputed (see, e.g., \cite{Kefayati2024}). 
It has meanwhile become {mostly} accepted that the laser-induced demagnetization of the FM layer leads to injection of a spin current from the FM layer into the NM heavy-metal layer, where the inverse spin Hall effect (ISHE) leads to the conversion of the spin current to a charge current pulse which, through electric dipole emission, causes the THz radiation.
However, two aspects continue to be debated. The first aspect is the source of the spin current. The spin current has been attributed to several different physical mechanisms, such as the generation of a spin-polarized current of non-thermal electrons in the superdiffusive transport regime  \cite{Battiato2010,Battiato2012,Kampfrath2013}, or to the generation of a thermal spin current through the spin Seebeck effect \cite{Seifert2018,Choi2015,Alekhin2017}. Another recently developed model, called $dM/dt$ model, attributes the spin current to spin pumping from the FM to the NM layer \cite{Beens2020,Lichtenberg2022}. Also, diffusive spin transport has been used to model the THz emission \cite{Dang2020}. {Recently, it was furthermore argued that the spin-to-charge conversion in the NM heavy-metal layer does not give an essential contribution to the THz emission \cite{Kefayati2024}.}

A second debated issue is whether the emitted THz electric field is proportional to the charge current $\bm{J}$ or to its time derivative, $\partial \bm{J} / \partial t$ (see e.g., Refs.\ \cite{Seifert2016,Huisman2015PRB,huisman2016NatNano,Nenno2019,Pettine2023,Varela2024,Kefayati2024}).
Obviously, the THz bandwidth depends strongly on whether the THz signal is due to the charge current or its derivative, as the latter will display a stronger time-dependence and thus a much wider bandwidth.  
 Taken together, the lacking understanding in the source of the spintronic THz emission, how it is generated and leads to the measured THz electric field hampers the development of efficient spintronic THz emitters as well as the precise extraction of typical spintronic quantities from the THz signal, such as the spin diffusion length \cite{Gorchon2022}.

Our aim here is to provide a quantitative theoretical description of spintronic THz emission.
We focus first on two aspects behind THz emission: the ultrafast dynamics of electrons in the emitter and the relation between the excited spin current and the generated THz electric field, which is experimentally well reported but not fully justified theoretically \cite{Nenno2019,Kefayati2024}. 
There {exists} namely an inconsistency between the experimental observations and theory based on Maxwell's equations, specifically, Jefimenko's equation \cite{Jefimenko1992}, which establishes that the emitted electric field should be proportional to the time derivative of the electric current. We solve this issue by taking into account the influence of the experimental setup and show that there is no contradiction, because the detecting system effectively integrates the THz signal in time. {Considering a Co(2 nm)/Pt(4 nm) bilayer as a typical spintronic THz emitter \cite{Wu2017,Qiu2018,Li2019,Chen2019,Dang2020},} 
we {then} use the superdiffusive (SD) spin transport theory \cite{Battiato2010,Battiato2012,Balaz2023} to describe the inner dynamics of electrons upon ultrafast laser excitation, and specifically show that the dependence of the spin Hall effect on the energy of the laser excited electron plays a role for the bandwidth of the THz emission.
{We further show that the duration of the excitation pulse strongly affects the detected THz signal. Lastly, we take into account  the thickness of the electro-optical detector crystal in the simulations to obtain realistic THz emission profiles.}

{In the following, we first introduce our theoretical modeling of THz emission in Sec.\ \ref{sec:theory}, followed by {computed} results for a typical Co/Pt emitter in Sec.\ \ref{sec:results}.}

\section{Model}
\label{sec:theory}

\subsection{THz electric field emission}
\label{sec:theory_E_emiss}

To describe the electric field $\bm{E}(\bm{r},t)$ generated {into three-dimensional space} from the charge current and the time-varying charge densities in the {metallic bilayer}, 
we employ Jefimenko's equation \cite{Jefimenko1992}
{\small
\begin{equation}
\begin{split}
    \bm{E}(\bm{r},t)&=\frac{1}{4\pi\epsilon_0}\int d \bm{r}'\left[\frac{(\bm{r}-\bm{r}')}{|\bm{r}-\bm{r}'|^3}\rho(\bm{r}',t_r)+\right.\\
    &\left.\frac{1}{c}\frac{(\bm{r}-\bm{r}')}{|\bm{r}-\bm{r}'|^2}\frac{\partial \rho(\bm{r}',t_r)}{\partial t}\right.
    \left.-\frac{1}{c^2}\frac{1}{|\bm{r}-\bm{r}'|}\frac{\partial \bm{J}(\bm{r}',t_r)}{\partial t}
    \right].
    \label{eq:Jefi4}
\end{split}
\end{equation}}
\noindent
In this expression, {which is the exact solution of the Maxwell equations,}  the sources of the THz emission are, respectively, the charge density $\rho$, its time derivative, $\partial \rho / \partial t$, and the time derivative of the charge current density $\partial \bm{J}/ \partial t$. The integral runs over the volume of the spintronic emitter, where $\bm{r}'$ represents the spatial element inside the emitter and $t_r$ is the retarded time $t_r=t-{|\bm{r}-\bm{r}'|}/{c}$. The retarded time takes into account the change in the source terms happening on timescales comparable to the emission timescales; {i.e.,} the speed of excited electrons moving in the spintronic emitter is of the order of nm/fs, which is not negligible with respect of the speed of light $c$ ($\approx$ 300 nm/fs). A short derivation of Eq.\ (\ref{eq:Jefi4}) is given in {Appendix} \ref{app:Jefi_eq}.

The three terms in Eq.~(\ref{eq:Jefi4}) have different decay properties: when the detector is  far away from the system (far field limit), the sign of the term $\bm{r}-\bm{r}'$ is well defined, and so the first term decays as inverse of the square of the distance while the second and third decay simply as the inverse of the distance. In {Appendix} \ref{app:Jefi_far} we discuss how  the contribution of the second term is zero due to conservation of charge, while the first term, by using the continuity equation $\partial \rho / \partial t + \bm{\nabla}\cdot \bm{J} =0$, becomes proportional to the current density. 
Comparing the decay properties, the third term that decays as the inverse of the distance will contribute the most in the far field. This observation is consistent with the THz electric field being proportional to $\partial \bm{J} / \partial t$, as has been stated in several recent investigations \cite{Nenno2019,Kefayati2024,Pettine2023,Varela2024}. 

{A few summarizing comments are appropriate at this point. Jefimenko's equation gives two contributions, a near-field contribution, the first term in Eq.\ (\ref{eq:Jefi4}), for which the THz field is proportional to the charge current $\bm{J}$ (as we show in Appendix \ref{app:Jefi_far}), and a far-field contribution proportional to $\partial \bm{J}/\partial t$.
This is consistent with previous work by Kaplan \cite{kaplan1998diffraction}, who solved the paraxial wave equation and found that, in the far-field limit, the traveling pulse becomes proportional to its time derivative.}

{On the other hand, several experimental works \cite{Seifert2016,Huisman2015PRB,huisman2016NatNano} used a different method to estimate the emitted THz field. In these works, a common element is the use of a one-dimensional wave equation to compute the electric field at the vacuum interface. As a result, they obtain a current-proportional field, $\bm{E} \propto \bm{J}$. 
This solution does not depend on the distance from the emitter, which is due to the choice of solving the wave equation in one dimension; i.e., one obtains a solution that does not decay in space. However, Jefimenko's equation gives that, far from the source in three dimensions the electric field decays proportional to the inverse of the distance, consistent with energy conservation. But in one dimension, there is no geometrical dilution of the emitted energy and, as such, the emitted wave does not decay with the distance.  For this reason, the solution proposed in the above works is incomplete, 
even though it could provide a correct estimate of the emitted near-field THz radiation, and we consider our approach based on the Jefimenko's equation, which is a solution in the whole space, more general.}

\subsection{Influence of detector on THz signal}
\label{sec:detector}

\begin{figure}[t!]
    \centering
\includegraphics[width=.9\linewidth]{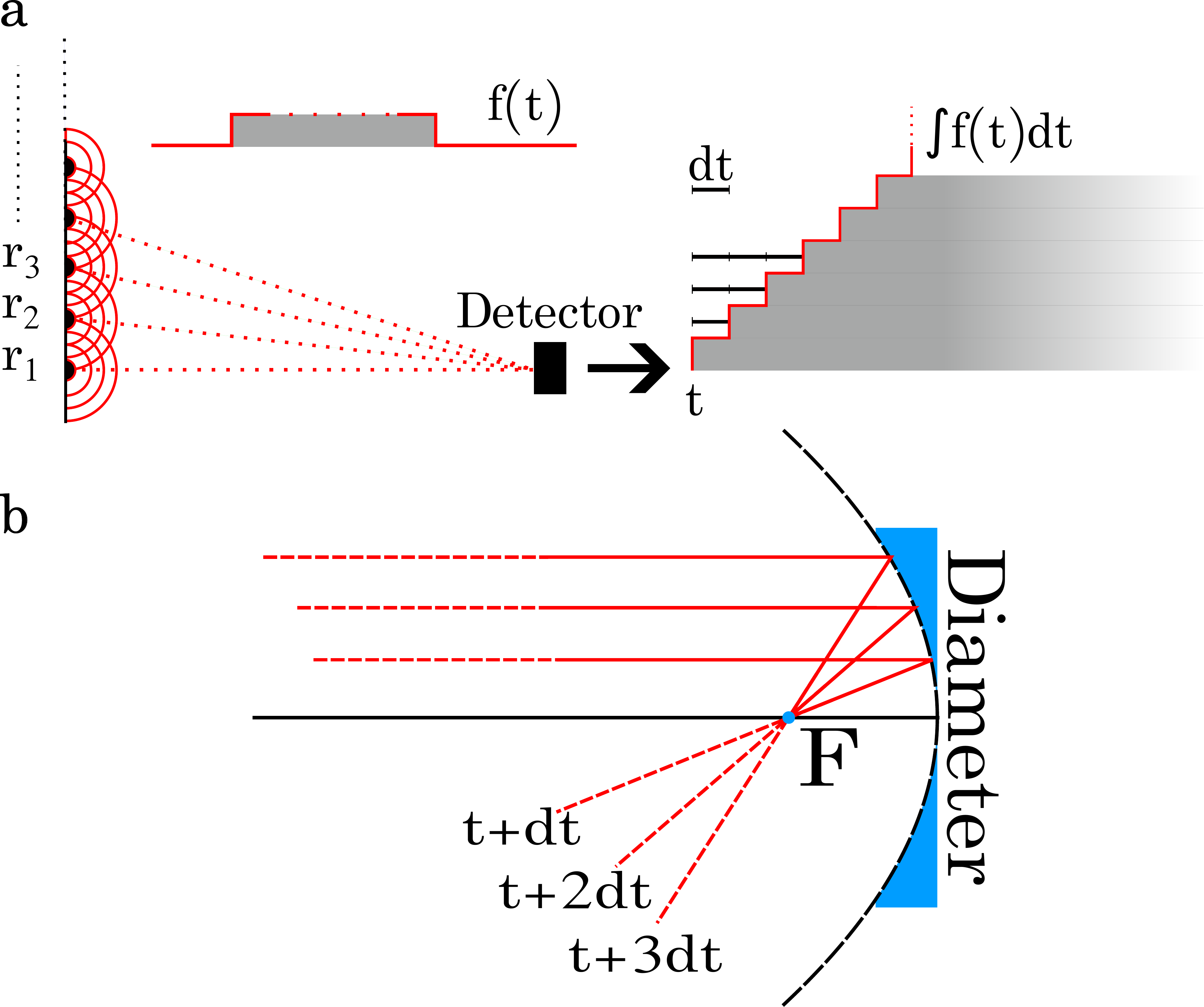}
    \caption{{Schematic representation of the integration of the THz signal over a surface.} ({a}) As the signal $f(t)$ is emitted from multiple source points $r_i$, these signals reach the detector at different times, where they are summed with a time delay proportional to 
    $dt$, resulting in a time integration of the signal. 
    ({b}) Focusing of the signal by the parabolic mirror: every point of the mirror acts as an emission point, the signal reaches the focal point $ F$ at a different time for every  emission point.}
    \label{fig:parab_M}
\end{figure}

To address the issue {of the proportionality of the THz signal}, we consider now
the influence of the presence of a macroscopic detector on the shape of the detected THz signal.
In the measurements, the radiation emitted from the sample is propagated through the experimental setup and collected into the detector with the use of {several} parabolic mirrors. {The presence of these mirrors, especially if not perfectly aligned, could introduce delays or distortions in the collected signal, however, the following conclusions hold true independently of other optical elements besides the last mirror.}
{Thus,} we can consider the {last} mirror, the one that focuses the radiation to the detector, effectively as a macroscopic emitter where every point of the mirror can be considered as an emission center. 
When a non-point-like source emits a signal, the signal emitted from every different point reaches the detector at different times. As a result the signal is effectively integrated in time, as sketched in Fig.~\ref{fig:parab_M}(a), resulting in an electric field which is proportional to the current in the spintronic device, as outlined below. 
Figure~\ref{fig:parab_M}(b) gives an intuitive picture on how the reflected radiation from every point of the mirror would reach the detector with a time delay due to the position of the different points of the mirror.

The detected electric field can thus be obtained by considering the far-field term in Eq.~(\ref{eq:Jefi4}) and integrating it over the surface of the mirror
\begin{equation}
    \bm{E}(t)\propto \int d \bm{s} \, \frac{\partial \bm{J}}{\partial t}(t_r) \, ,
    \label{eq:E_mirror}
\end{equation}
where $d{\bm{s}}$ is the element of surface area. 
The time dependence of the current with respect of the emission point is contained in the retarded time $t_r$, where the distance from the focal point of the mirror (where the detector is located) causes different points on the mirror to correspond to different values of $t_r$.
The straightforward relation between the space and time coordinates allows us to rewrite the integral in Eq.\ (\ref{eq:E_mirror}) as an integral over $dt$, for technical details, see {Appendix}\ \ref{app:cent_mirr}. Therefore, after the integration, the measured time-dependent electric field results proportional to the charge current, i.e.,  $\bm{E}(t)\propto \bm{J}(t)$, even in the 
far field. 
{The solution to the apparent contradiction of an emitted electric field proportional to the current derivative or the current itself lies thus in the way the emitted field is collected into the detector.}

It should additionally be mentioned that, for the examined detection scheme, the parabolic mirrors used to collect the radiation into the detector are off-axis parabolic mirrors. These mirrors are obtained as a section of a three-dimensional paraboloid, so that the radiation is incident on the mirror and is reflected into the detector with an angle.  In the above, we considered for simplicity a \textit{centered} mirror, where the direction of incident and reflected radiation is parallel to the optical axis of the mirror. This could obviously not be a suitable experimental setup, as the detector would screen the mirror from the radiation emitted by the device, hence the need for off-axis mirrors.
A longer derivation given in {Appendix} \ref{app:integ} shows that this approximation does not change the results qualitatively.

\subsection{The superdiffusive spin-transport model}
\label{sec:theory_superdiff}

We use here the SD spin-transport theory \cite{Battiato2010,Battiato2012,Lu2020,Balaz2023}  to model the laser-induced ultrafast spin current in a selected FM/NM system.  The laser pulse excites spin-up and spin-down electrons in the FM layer of the spintronic emitter and, since the two spin channels have different transport properties (i.e., different velocities and lifetimes of electrons), the excited current that is injected from the FM to the NM layer is spin polarized. The spin current flux into the NM is the amount of demagnetization of the FM layer,
{$\partial M/ \partial t$}.

The spintronic emitter is described as a quasi-1D system along the $z$ direction, normal to the heterostructure. {This approximation is justified by the nanometer thickness of the emitter, compared to the micrometer-size spot of the pump laser.}
The density of excited electrons $n_\sigma(\epsilon,z,t)$, with $\sigma$ being the electron spin, $\epsilon$ its energy and $(z,t)$ the space and time coordinates, is described by the {transport} equation 
\begin{equation}
    \frac{\partial n_   \sigma}{\partial t}+\frac{n_   \sigma}{\tau_   \sigma}=\left(-\frac{\partial}{\partial t}\hat\phi +\hat I \right)S^{\mathrm{eff}}_   \sigma,
    \label{eq:superdifussion}
\end{equation}
where, for simplicity, we omitted the dependence on the variables $(\epsilon,z,t)$. Here, $\tau_\sigma=\tau_\sigma(\epsilon,z)$ is a hot electron lifetime that depends on the energy and spin of electrons and the material we consider. 
The operator $\hat \phi$ describes the electron flux in the system, and contains information about scattering events, interactions among electrons, between electrons and the lattice, spin flip, lifetime and velocities of the electrons. The term $\hat I$ is the identity operator, while $S^{\mathrm{eff}}_\sigma$ represents the effective spin source, consisting of both the contributions from the laser-excited electrons $S^\mathrm{ext}_\sigma$, which is modeled as a Gaussian shaped pulse, and the ones excited through scattering events $S^\mathrm{p}_\sigma$ so that $S^\mathrm{eff}_{\sigma}=S^\mathrm{ext}_{\sigma}+S^\mathrm{p}_{\sigma}$.  
Solving Eq.~(\ref{eq:superdifussion}) for a chosen heterostructure we obtain the non-thermal electron density $n_\sigma(\epsilon,z,t)$ and the spin-current density $j_s(z,t)$ from which we compute the charge current and THz emission.

The SD spin transport theory is a materials' specific theory through the hot electron lifetimes and velocities. We therefore consider specifically a Co(2 nm)/Pt(4 nm) system as representative of a typical THz emitter \cite{Wu2017,Qiu2018,Li2019,Chen2019,Dang2020}. The spin-dependent hot electron lifetimes and velocities are taken from first-principles calculations \cite{Zhukov2006,Nechaev2010}. We furthermore assume excitation of the THz layer by an 1.5-eV pump laser, as used in most experiments, and investigate the influence of the pump pulse duration.

\subsection{Spin-to-charge conversion}
\label{sec:theory_s2c}

When the spin current is injected in the NM layer, it is converted via the ISHE \cite{Dyakonov1971_current,Hirsch1999,Kampfrath2013} into a transient charge current as $J_c(z,t) =  \frac {2 e}{\hbar}\theta_\mathrm{SH} J_s(z,t)$, where $\theta_\mathrm{SH}$ is the 
spin Hall angle. $\theta_\mathrm{SH}$ is defined in terms of the spin Hall conductivity $\sigma_\mathrm{SH}$ and the normal electrical conductivity $\sigma_e$ as $\theta_\mathrm{SH}=2\sigma_\mathrm{SH}/\sigma_\mathrm{e}$ and it represents the efficiency of the spin-to-charge conversion process. 

The rate of spin-to-charge conversion depends on $\theta_\mathrm{SH}$ and, hence, on the spin Hall conductivity, which is a quantity that is usually energy dependent \cite{Salemi2022}.
In our simulations, the NM material is Pt, which has a strong energy-dependent spin Hall conductivity \cite{Salemi2022}. In particular $\sigma_\mathrm{SH}$ for Pt is large for electrons close to the Fermi level, but it decreases significantly for energies higher than 0.5~eV, meaning that excited electrons with energy higher than this value will give negligible contributions to the charge current. However, high energy electrons can still contribute to the emission via scattering processes. When electrons scatter, they {lose energy and} decay into lower energy levels and populate the energy channels which can give significant contributions to the THz emission.

{To include the energy dependence of the spin-charge conversion we proceed as follows: When}
solving Eq.\ (\ref{eq:superdifussion}) we obtain the energy-dependent electron density $n_\sigma(\epsilon,z,t)$, where the energy $\epsilon$ assumes values from 0 to 1.5~eV in 12 discrete energy levels. As only the levels of energy not higher than 0.5~eV contribute to the emission, only the first three energy levels can efficiently be converted into a charge current through ISHE. 
We suppose {that} every electron can decay from its energy level to the energy below with a decay time of $\tau_\epsilon$. For simplicity, we do not consider the case where electrons can decay multiple levels at once, but must go through all the intermediate levels if they decay into a level which is not immediately below in energy. Additionally, we {assume} that spin is conserved in the decay, i.e.\ no spin flip happens while the electron loses energy.  Therefore, after the decay, the new electron density $n_\mathrm{dec}$ 
is described by the following equation
\begin{equation}
\begin{split}
    n_\mathrm{dec}(\epsilon,t+\delta t)=\Big[ n(\epsilon,t+\delta t)+n_\mathrm{dec}(\epsilon+1,t+\delta t) \\ +e^{-\frac{\delta t}{\tau_\epsilon}}\left(n(\epsilon,t)+n_\mathrm{dec}(\epsilon+1,t)\right)\Big](1-e^{-\frac{\delta t}{\tau_\epsilon}}),
\end{split}
\label{OPT-eq:decay}
\end{equation}
where the spin index has been omitted, for the sake of simplicity. 

We point out that, for every energy level, $n_{dec}$ is composed of contributions from the original excited density $n$ (electrons of that level that didn't decay) and from the electrons that decayed from higher-lying energy levels. Finally, we use the three lowest energy levels of the new computed electron density to compute the charge current that will then give rise to the THz emission.

\begin{figure}[htbp]
    \centering
    \includegraphics[width=.9\linewidth]{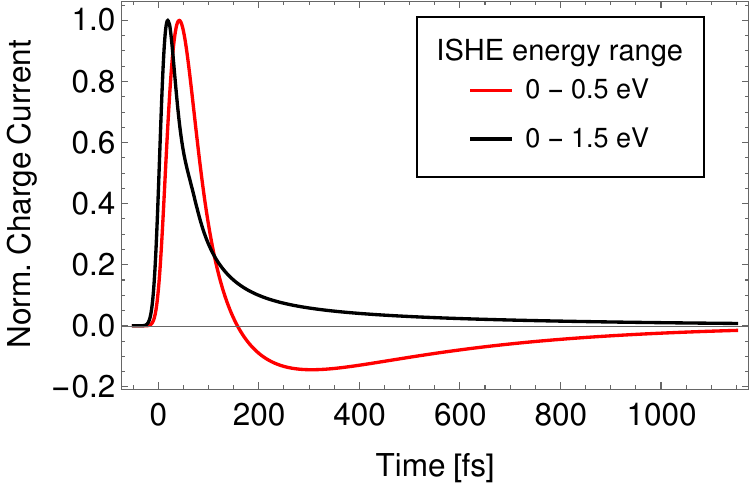}
    \caption{{Influence of the energy dependence of the ISHE on the transient charge current.} The black line gives the computed charge signal when all excited electrons contribute to the spin-charge conversion and the red curve when only the spin-polarized hot electrons with energies up to 0.5 eV contribute. A pump pulse of 20 fs was used in the calculations.
    }
    \label{fig:cascade}
\end{figure}

Figure \ref{fig:cascade} shows the influence of the energy dependence of the spin-to-charge conversion due to the ISHE on the computed charge current, for the Co(2 nm)/Pt(4 nm) bilayer excited with a 20-fs laser pulse. {We note that, if the spin Hall angle is kept uniform in the $0-1.5$~eV range, then the charge current would simply be proportional to the spin current which, in the superdiffusion theory, is computed through the spin flux in the sample \cite{Battiato2010,Battiato2012,Lu2020}. However, when using an energy-dependent spin Hall angle, the exact proportionality between  charge current and spin current is lost. In particular, we compute the charge current by using the continuity equation in conjunction with the new electron density $n_{dec}$ computed in Eq.~\eqref{OPT-eq:decay}, restricted to the lower energy levels. This results in the change in the shape of the charge current with respect to the spin current shown in Fig.\ \ref{fig:cascade}. Moreover,} the charge current pulse peaks at a later time, as expected, since excited spin-majority electrons need to decay to lower energy levels to contribute to the charge current. The influence of later arriving spin-minority electrons becomes in addition more pronounced. 

\section{Results}
\label{sec:results}

\begin{figure}[htbp]
    \centering
    \includegraphics[width=.85\linewidth]{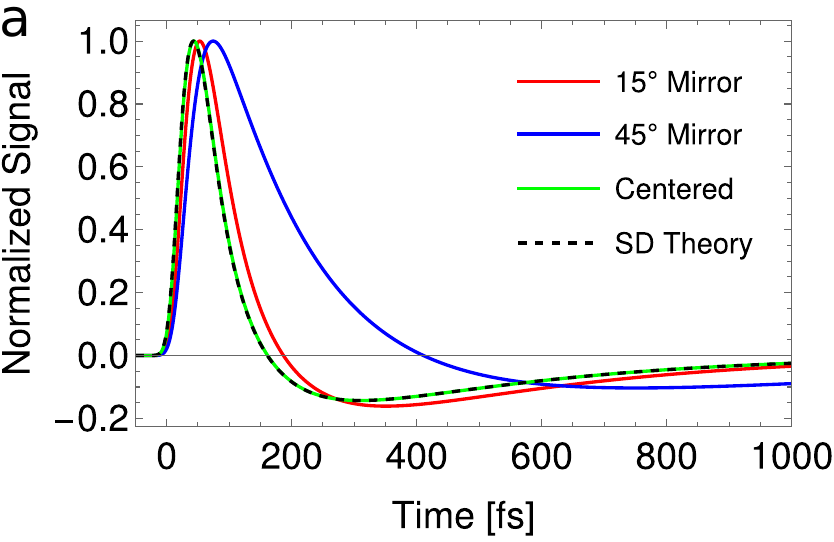}
    \includegraphics[width=.85\linewidth]{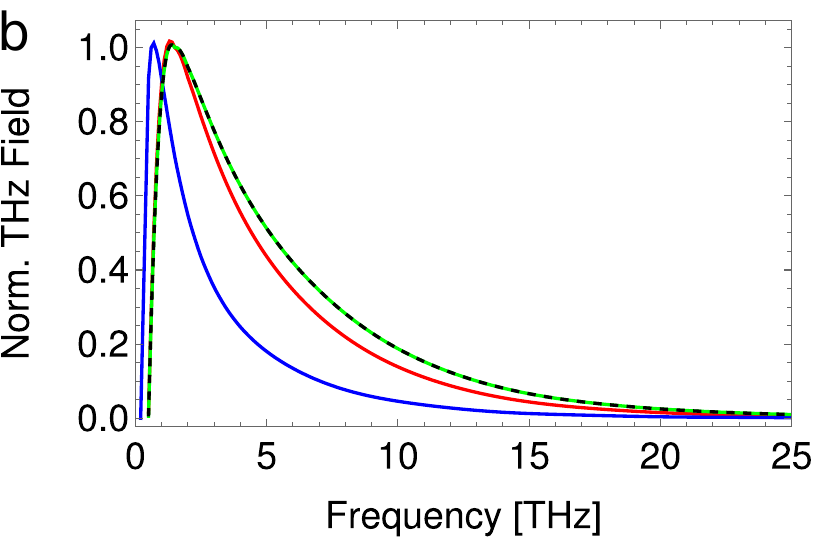}
    \caption{{Influence of detector mirror on the THz signal of the Co/Pt emitter.} ({a}) Transient THz electric field after the integration over the surface of the mirror for 15$^{\circ}$ (red), 45$^{\circ}$ (blue) off-axis and centered (green curve) mirrors compared with the {charge} current obtained from the SD model (dashed black curve).
    ({b}) Bandwidth of the THz signals obtained after integrating over the surface of the mirror (same color legend as under ({a}), compared with the bandwidth obtained from the calculated charge current (dashed black line). 
   {The curves have been normalized to the same peak value for easier visualization.}
    }
    \label{fig:off-axis_signal}
\end{figure}

\begin{figure*}[htbp]
    \centering
    \includegraphics[width=.75\linewidth]{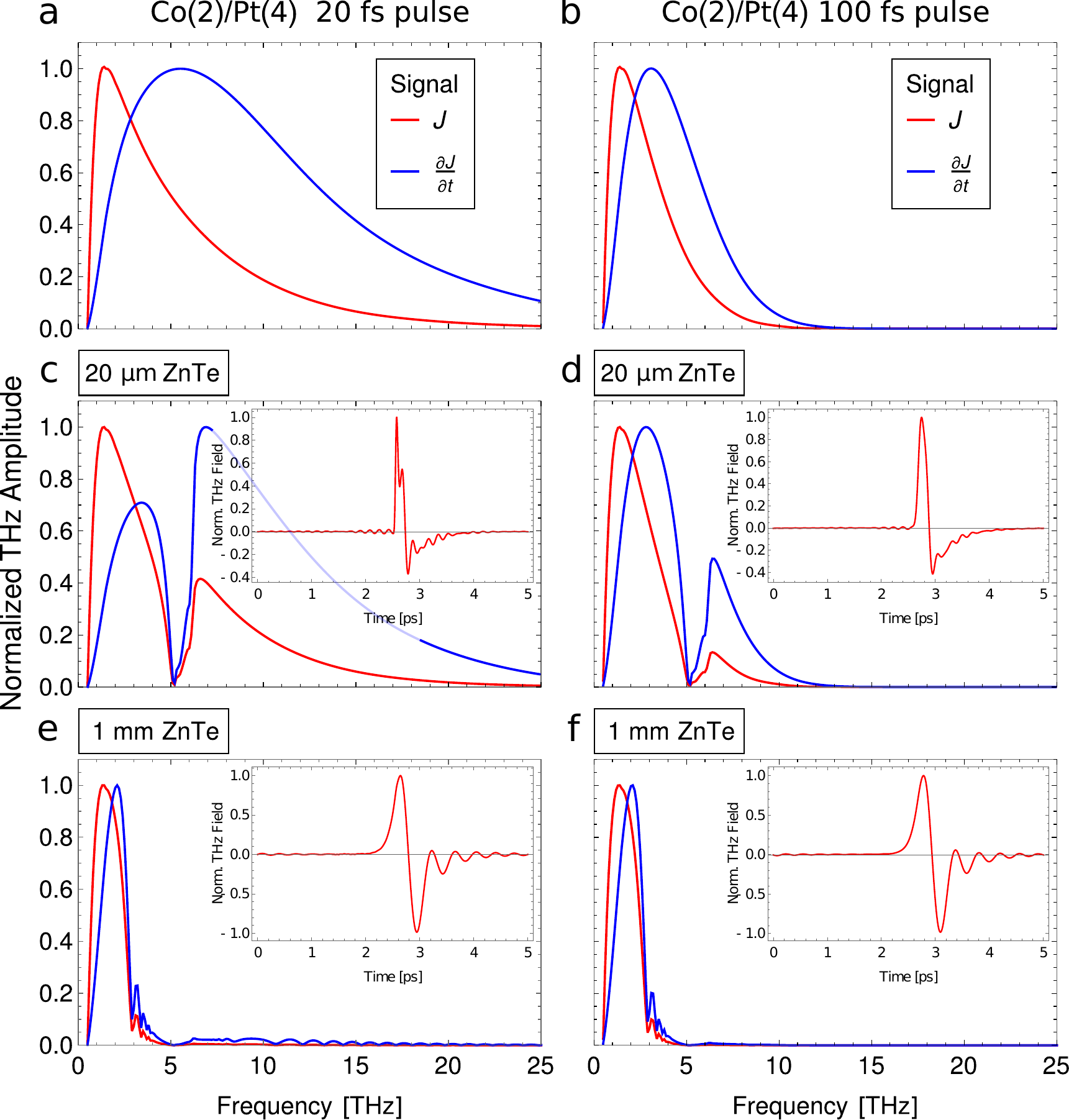}
    \caption{{Spintronic THz spectra computed for a Co/Pt emitter.} The THz bandwidth generated from the charge current (red) and the charge-current derivative (blue) are compared for short (left panels) and long (right panels) pump laser pulses. (a) Simulated signals generated by the charge-current derivative and by the charge current for a 20-fs laser pulse, and ({b}) for a 100-fs long pump pulse. 
    (c) and (d) Simulated THz signals convoluted with the response function of a 20 $\mathrm{\mu}$m thick ZnTe electro-optical crystal for short ({c}) and long pulses ({d}). ({e}) and ({f})  Simulated THz signals convoluted with the response function of a 1 mm thick ZnTe crystal for short ({e}) and long pulses ({f}).
    The insets in panels ({c})-({f}) show the time trace of the THz electric field  generated by the charge current $J$.}
    \label{fig:jdj_results}
\end{figure*}

\subsection{Detector orientation}

{We start with comparing} 
the {charge} current 
in the Co/Pt spintronic device with the {collected THz electric field}
for different mirror configurations, assuming a 20-fs laser pulse.
We consider here three cases: a centered mirror, a 15$^{\circ}$, and a 45$^{\circ}$ off-axis parabolic mirror, where the angle refers to the angle between the incident radiation and the optical axis of the mirror, i.e.\ the axis that connects the center of the mirror with the focus of the parabola, see {Appendix} \ref{app:off_mirr} for details. All the mirrors have a diameter of 50~mm and a focal distance of 100~mm. 

Figure~\ref{fig:off-axis_signal}(a) shows the comparison between the transient charge current profile in the Co/Pt spintronic device (black dashed line) with the collected electric field signal for the considered mirrors. We can see that, {after employing Jefimenko's equation (\ref{eq:Jefi4}) to obtain the far field $E(t)$,} the integration over the centered mirror (green line) perfectly superimposes with the current profile from the emitter. {Thus, the focusing mirror effectively converts the THz signal proportional to $\partial J/ \partial t$, to an $E(t) \propto J(t)$ signal.}  For the off-axis mirrors, we see that the integration over the 15$^{\circ}$ mirror produces a signal really similar to the original current profile, with a slight stretch of the profile in time, while,  the 45$^{\circ}$ mirror induces a significantly {larger time stretch}.

The 
mirror configuration has interesting effects on the measured bandwidth, too. Figure~\ref{fig:off-axis_signal}(b) shows, in Fourier space, the resulting electric field as computed from the original current and from the THz signal that has been integrated over the surface of the mirrors. As for the previous case, the signal integrated on the centered mirror provides an exact superposition with the profile of the original current, while the 15$^{\circ}$ signal appears to be slightly reduced in bandwidth 
and the signal for the 45$^{\circ}$ mirror has a 
bandwidth that is more reduced with respect to that from the original source. 

\subsection{Simulated THz bandwidth}

We now consider the influence of the laser pulse duration and of the electro-optical sampling with a non-linear ZnTe crystal.
Figure~\ref{fig:jdj_results} shows the computed THz bandwidth for the Co(2 nm)/Pt(4 nm) system. 
For comparison, we show the THz bandwidths computed for the $E$ field proportional to the current or to the current derivative.
For sake of simplicity, 
we consider the case of a centered mirror. 
The received THz signal is then exactly proportional to the charge current and not distorted as shown in Fig.~\ref{fig:off-axis_signal}, so that a direct comparison with the current-derivative signal, computed in absence of mirrors (and hence not distorted), is straightforward.

Furthermore, we take into account the effects of different exciting laser pulse lengths (1.5~eV, 20~fs and 100~fs) as well as the response function of a ZnTe crystal commonly used used for the THz detection \cite{wu_comparison_2014,casalbuoni_numerical_2008,gallot1999electro}.
A number of very interesting consideration can be extracted from our results. 

First, let us consider only the predicted THz emission profile, without taking into account the response function of the detecting crystal, in Figs.~\ref{fig:jdj_results}(a) and \ref{fig:jdj_results}(b).
For short exciting laser pulses, the difference between the THz signal induced by the current or by the current derivative is striking; ${E} \propto \partial {J}/ \partial t $ 
results in a very broad emission, of the order of 30 THz, while 
${E} \propto {J} $
gives a much narrower emission, of the order of $5-10$ THz. However, when laser pulses with longer pulse width ($\approx$ 100~fs) are used to excite the system, the differences between the two emission profiles appear much less pronounced, with a more similar bandwidth. Therefore, when comparing THz signals, particular attention must be paid to the used pulse width, since long exciting pulses would make it particularly challenging to distinguish the two emission mechanisms.

Second, the response function of the detection crystal plays a fundamental role in the detection.  Figures~\ref{fig:jdj_results}(c)-\ref{fig:jdj_results}(d) show the predicted THz emission profiles convoluted with the response function of a ZnTe crystal \cite{wu_comparison_2014}; see Appendix Sec.~\ref{app:response} for details regarding the derivation of the response function. We consider two cases, a thin (20~$\mu$m) and thick (1~mm) ZnTe crystal. The use of a crystal of a certain thickness introduces an additional filter to the detected THz frequency spectrum. If the crystal is thin enough, the detected spectrum is not considerably reduced, as we can see in the case of the 20~$\mu$m crystal:  aside from a sharp dip at $\approx$~5~THz, the detected bandwidth is still large enough
{so that emission profiles due to the charge current or its derivative can be distinguished.}

For short exciting pulses (20 fs) the ${E} \propto {J}$ and ${E} \propto \partial {J}/ \partial t$ signals are remarkably different, but this difference is less evident for long pump pulses instead. 
The case of the thick electro-optical crystal is also very interesting. As the thickness of the crystal increases, the frequency window of the detector is dramatically reduced (see {Appendix} Fig.~\ref{sfig:ZnTe}), therefore reducing both the signal computed from the current and the current derivative, for both short and long excitation pulses, leading to THz spectra being almost indistinguishable in practice.

{The insets in Figs.\ \ref{fig:jdj_results}(c)-\ref{fig:jdj_results}(f) show the time-domain electric signal due to the charge current $J$ and include the influence of the thickness of the ZnTe crystal. The THz time traces exhibit a characteristic bipolar shape comparable to THz pulses observed in experiments
\cite{Seifert2018,Qiu2018,Chen2019,Dang2020,Kolejak2024}. Analogous to the observations made for the THz bandwidth, the time traces for long and short pump pulses are practically indistinguishable when a 1-mm thick electro-optical crystal is used for detection.}

\section {Discussion and Conclusions}

Several aspects of spintronic THz emission have been debated recently. First, the source of the emission was disputed. Magnetic dipole emission due to a rapid change of magnetization {($E \propto \partial^2 M / \partial t^2$)} was first proposed as source of THz emission \cite{Beaurepaire2004}. Such source of THz emission could occur in single thin ferromagnetic films, but it has been shown that in FM/NM bilayers the electric dipole emission is responsible for the much stronger THz emission and {moreover consistent with spin-charge conversion due to the ISHE}
\cite{Kampfrath2013,Zhang2020}.
 {Second, different proportionalities of the emitted THz field have been proposed, being proportional to either the charge current or the time-derivative of the charge current \cite{Seifert2016,Seifert2018,Nenno2019,Pettine2023,Varela2024,Kefayati2024}. Although the proportionality to $\partial {J} / \partial t$ is consistent with Maxwell theory, our investigation  of the influence of the parabolic mirror shows that this detection method effectively leads to a time integration of the signal, making it proportional to the charge current.}
 {The measured far-field THz signal acquires then a current-field proportionality, which is similar to the proportionality obtained from a one-dimensional wave-equation \cite{Huisman2015PRB,huisman2016NatNano}, but obviously for entirely different reasons.}
{It deserves to be mentioned, too, that apart from the ISHE, operative in the bulk metallic NM layer, THz emission due to a local spin-charge conversion as the inverse Rashba-Edelstein effect was proposed for systems with a Rashba surface state, such as a FM layer covered with a thin Ag/Bi layer \cite{Jungfleisch2018,Zhou2018}, or with ultrathin Bi$_{1+x}$Sb$_x$ \cite{Rongione2023}.}
  
{We further find that the experimental configuration (detector crystal and mirror configuration) as well as the duration of the laser pump, 
can affect the profile of the recorded THz signal substantially.}
{In essence, we obtain that, when the exciting pulses are relatively long ($\approx 50-100$~fs) or the detector crystal is quite thick ($\approx$ 1~mm), the THz bandwidth becomes reduced. The emitted THz spectrum for the two cases (${E} \propto {J}$ or $\propto \partial {J}/ \partial t$) are then hardly distinguishable. This may provide an explanation for why the debate between current- and current-derivative THz emission had not found a conclusion yet.}
{We further conclude from our simulations that, to distinguish clearly between a current and a current-derivative THz signal, thin electro-optic crystals, i.e., less than 50~$\mu$m, should be used for detection in combination with short pump laser pulses, of the order of 20~fs.}

{Lastly, the origin of the spin current generated by the pump laser pulse has been debated \cite{Kampfrath2013,Choi2015,Alekhin2017,Seifert2018,Nenno2019,Beens2020,Lichtenberg2022,Kefayati2024}.
The different proposed mechanisms for spin current generation, such as, superdiffusion \cite{Battiato2010,Battiato2012,Balaz2023}, spin Seebeck effect \cite{Choi2015,Alekhin2017,Seifert2018}, and spin current due to demagnetization \cite{Beens2020,Lichtenberg2022}, could be difficult to distinguish in practice. In superdiffusion, the excited spin-polarized electrons are non-thermal, whereas in the spin Seebeck effect and in spin pumping the electrons are considered to have thermalized. Optically induced superdiffusive spin currents have been detected in various metallic heterostructures \cite{Melnikov2011,Rudolf2012,Hofherr2017,Gupta2023},  but, in early measurements of spintronic THz emission it appeared that the superdiffusive spin current was faster than the measured charge current \cite{Kampfrath2013}. 
Taking however the integrating effect of the collecting mirror into account, quantitative modeling of the THz signal with bandwidths that are overall in agreement  with measurements is obtained. 
Also, as pointed out in Sec.\ \ref{sec:theory_s2c} the SHE has often been assumed to be independent of the electron energy, but a more precise description can be obtained when its energy dependence is included. 

Our investigation particularly emphasizes that to explain measured THz emission, several processes need to be modeled, ranging from the real-time excitation and injection of non-equilibrium spin-polarized electrons to their decay due to scattering, spin-charge conversion by an energy-dependent ISHE, and simultaneously occurring electric dipole emission, to the influence of the detection setup on the far-field electric signal.  Taking these into account a quantitative explanation of spintronic THz emission can be  achieved. 
}

\appendix

\counterwithin{table}{section} 

 \begin{widetext}
\section{Electric field as a function of its sources}
\label{app:E(J)}
\subsection{Jefimenko's equation for the electric field} 
\label{app:Jefi_eq}
Here we provide a brief derivation of the electric field $\bm{E}(\bm{r},t)$ in terms of its sources, the charge current density $\bm{J}$ and the charge density $\rho$. To this end, we consider the retarded potentials
\begin{equation}
    \phi(\bm{r},t)=\frac{1}{4\pi\epsilon_0}\int d\bm{r}'\;\frac{\rho(\bm{r}',t_r)}{|\bm{r}-\bm{r}'|} \qquad \textrm{and}\qquad
    \bm{A}(\bm{r},t)=\frac{\mu_0}{4\pi}\int d\bm{r}'\; \frac{\bm{J}(\bm{r}',t_r)}{|\bm{r}-\bm{r}'|} ,
\end{equation}
where $\bm{r}'$ represents the coordinate inside the domain of integration, $d\mathbf{r}'$ is the volume element and $t_r=t-{|\bm{r}-\bm{r}'|}/{c}$ is the retarded time. By substituting the expressions for the potentials in the general expression for the electric field, $\bm{E}(\bm{r},t)=-\nabla\phi(\bm{r},t)-{\partial \bm{A}(\bm{r},t)}/{\partial t}$ we obtain
\begin{equation}
\begin{split}
    \bm{E}(\bm{r},t)=&-\nabla\left(\frac{1}{4\pi\epsilon_0}\int d\bm{r}'\;\frac{\rho(\bm{r}',t_r)}{|\bm{r}-\bm{r}'|}\right)-\frac{\partial }{\partial t}\left(\frac{\mu_0}{4\pi}\int d\bm{r}'\; \frac{\bm{J}(\bm{r}',t_r)}{|\bm{r}-\bm{r}'|}\right)\\
    =&\frac{1}{4\pi\epsilon_0}\left[-\nabla\int d\bm{r}'\;\frac{\rho(\bm{r}',t_r)}{|\bm{r}-\bm{r}'|}
    -\mu_0\epsilon_0\frac{\partial }{\partial t}\int d\bm{r}'\; \frac{\bm{J}(\bm{r}',t_r)}{|\bm{r}-\bm{r}'|}    
    \right].
\end{split}
\end{equation}
Note that $\nabla=\nabla_{\bm{r}}$ and both $\rho$ and $\bm{J}$ depend on $\bm{r}$ implicitly through $t_r$, and also $\partial/\partial t=\partial/\partial t_r$, so that
\begin{equation}   \nabla\rho(\bm{r}',t_r)=\frac{\partial \rho}{\partial t_r}\frac{\partial t_r}{\partial \bm{r}}=\left(-\frac{1}{c}\frac{(\bm{r}-\bm{r}')}{|\bm{r}-\bm{r}'|}\right)\frac{\partial \rho}{ \partial t_r}\qquad \textrm{and}\qquad
        \nabla\left(\frac{1}{|\bm{r}-\bm{r}'|}\right)\rho=-\frac{(\bm{r}-\bm{r}')}{|\bm{r}-\bm{r}'|}\rho,
\end{equation}
and this leads to
\begin{equation}
    \bm{E}(\bm{r},t)=\frac{1}{4\pi\epsilon_0}\int d\bm{r}'
    \left[\frac{(\bm{r}-\bm{r}')}{|\bm{r}-\bm{r}'|^3}\,\rho(\bm{r}',t_r)
    +\frac{1}{c}\frac{(\bm{r}-\bm{r}')}{|\mathbf{r}-\bm{r}'|^2} \, \frac{\partial \rho(\bm{r}',t_r)}{\partial t}
    -\frac{1}{c^2}\frac{1}{|\bm{r}-\bm{r}'|} \, \frac{\partial \bm{J}(\bm{r}',t_r)}{\partial t}\right] ,
    \label{Seq:Jefi4}
\end{equation}
which is the well-known Jefimenko equation for the {generated} electric field \cite{Jefimenko1992}.
\end{widetext}

\subsection{Contributions from Jefimenko's equation}
\label{app:Jefi_far}

The last two terms of Eq.~\eqref{Seq:Jefi4} decay as the inverse of the distance distance $|\mathbf{r}-\mathbf{r}'|$, thus they could be detected experimentally at long distances. The last term has already been discussed in the main text, so we focus our attention to the second one.

We show here that the contribution of the second term to the measured electric field, in the system described in the main text, is zero. 
First, we point out that the vectors $\bm{r}$ and $\bm{r}'$ represent, respectively, the position of the observer (detector) and the position of the emission center in the sample. As the distance between the detector and the sample is at least of the order of tens of cm, and as the dimensions of the emitter vary from the $\mu$m to the mm range, we can safely approximate $\bm{r}-\bm{r}'$ as $\bm{r}$. So that the contribution from the second term to the electric field, {denoted as} $\bm{E}^{(2)}$, becomes
\begin{equation}
    \bm{E}^{(2)}(\bm{r},t) \approx 
    \frac{1}{4\pi\epsilon_0} \frac{1}{c} \frac{\hat{\bm{r}} }{r}
    \int d\bm{r}'\frac{\partial \rho(\bm{r}',t_r)}{\partial t},
    \label{seq:second_term}
\end{equation}
where $\hat{\bm{r}}$ is the unitary vector in the direction of $\bm{r}$. The integral in Eq.\ \eqref{seq:second_term} is a volume integral of the time derivative of a charge density, i.e.\ it represents {the} total flux of charges in or out the volume of integration. As there are no sources of charge or any sink, the charge is conserved in the sample, thus the integral in Eq.\ \eqref{seq:second_term} must be zero, and the only far-field term contributing to the electric field is the third one, {as} has been {mentioned} 
in the main text. 

For the sake of completeness, we also study the behavior of the first term in Eq.\ \eqref{Seq:Jefi4}, which decays as the square of the distance. Here we show that it can be expressed in terms of the current density $\bm{J}$. Using the continuity equation, $\frac{\partial \rho}{\partial t}+\nabla\cdot\bm{J} =0$, we can rewrite the first term of Jefimenko's equation as 
\begin{equation}
   \frac{1}{4\pi\epsilon_0} \!\int \!\!d\bm{r}'
    \frac{(\bm{r}-\bm{r}')}{|\bm{r}-\bm{r}'|^3}\,\rho(\bm{r}',t_r)\approx    \frac{1}{4\pi\epsilon_0}
    \frac{\hat{\bm{r}} }{r^2}\!
    \int \! d\bm{r}'dt \nabla\cdot\bm{J}( \bm{r}',t_r)
    \, .
    \label{seq:first_term}
\end{equation}

We now focus on the volume integral. We consider a cuboid of dimensions ($\Delta x,\Delta y,\Delta z$) as the volume of integration that represents our sample and we use the flux theorem to compute the integral. As $\bm{J}$ is not uniform through the volume of integration, 
the integral is the sum of the fluxes of $\bm{J}$ at every element $z'$ (for simplicity we consider $\bm{J}$ uniform along $\hat{\bm{x}}$ and we neglect the $x$-dependence in the following)
\begin{equation}
    \int d \bm{r}' \nabla\cdot\left.\bm{J}(\bm{r}',t_r)\right.= \sum_{{z}'} S_y^{{z}'} [\bm{J}(y_f,z',t_r)-\bm{J}(y_i,z',t_r)] \, .
\end{equation}
$y_f$ and $y_i$ are the integration boundaries along $\hat{\bm{y}}$ and  $S_y^{{z}'}$ is the surface of the cuboid at the element $z'$ that is perpendicular to the current density $\bm{J}$.
If the dimension $\Delta y$ of the volume of integration is sufficiently small, we can rewrite the previous equation in term of the $y$-derivative of $\bm{J}$
\begin{equation}
    \int d \bm{r}' \nabla\cdot\left.\bm{J}(\bm{r}',t_r)\right.= \sum_{{z}'} S_y^{{z}'} \Delta y\frac{\partial \bm{J}(y,z',t_r)}{\partial y} .
\end{equation}
Additionally, we can rewrite the $y$-derivative in terms of the time derivative
\begin{equation}
    \frac{\partial }{\partial y}=\frac{\partial }{\partial t}\frac{\partial t}{\partial y}=\frac{1}{v_e}\frac{\partial }{\partial t} \, ,
\end{equation}
with $v_e$ the electron velocity in the material. Although electrons of different energies have different velocities, it is straightforward to generalize the equation above to the case of electrons of different velocities, but we {omit} it here for the sake of brevity.
We can now write
\begin{equation}
    \int d \bm{r}' \nabla\cdot\left.\bm{J}(\bm{r}',t_r)\right.=\frac{\Delta y}{v_e} \sum_{{z}'} S_y^{{z}'} \frac{\partial \bm{J}(z',t_r)}{\partial t} \, ,
\end{equation}
where for simplicity we removed the dependence on $y$. Finally, we can sum over $z'$, rewriting Eq.\ \eqref{seq:first_term} as
\begin{equation}
   \frac{1}{4\pi\epsilon_0}\int d\bm{r}'
    \frac{(\bm{r}-\bm{r}')}{|\bm{r}-\bm{r}'|^3}\,\rho(\bm{r}',t_r)\approx    \frac{1}{4\pi\epsilon_0}
    \frac{\hat{\bm{r}} }{r^2}\frac{\Delta V}{v_e}
    \bm{J}(t_r) \,
    \label{seq:nearF}
\end{equation}
where $\Delta y$ and the sum over $z'$ of $S_{y}^{z'}$ have been included in $\Delta V$, and {the} time derivative of the current has been integrated in time. 

To summarize, we showed that the second term of Jefimenko's equation, which is a far-field term, does not contribute to the observed signal due to conservation of charge, while the first term is directly proportional to the current density but is a near-field term, and hence decays much faster than the far-field terms.

\section{Integration over the surface of the mirror}
\label{app:integ}
\subsection{Centered Mirror}
\label{app:cent_mirr}

Here, we integrate the far-field term of Jefimenko's equation over the surface of a parabolic mirror to reproduce the detected THz signal. 

In a realistic setup, a number of mirrors and other optic elements are used to redirect and focus the radiation on the detector, and only the last mirror focuses the radiation on it. In the following derivation we consider only the last mirror, and we assume that the experimental setup focuses and redirects the radiation in such a way that it can be considered as a packet of many collinear beams that are all incident on the parabolic mirror with the same angle with respect to the symmetry axis of the mirror. 

For simplicity, we first consider the case of a centered mirror, where the emitted beams are parallel to the symmetry axis of the mirror.
The expression for the THz electric field emitted due to the current at the position $\bm{r}'$ and measured at position $\bm{r}$ and time $t$ is
\begin{equation}
    \bm{E}(\bm{r},t)\approx \int d\bm{r}' \frac{1}{|\bm{r}-\bm{r}'|}\frac{\partial \bm{J}}{\partial t}(\bm{r}',t_r).
\end{equation}
As the radiation emitted from the sample is reflected, we can consider the mirror as the new emitting system and integrate over the surface of the mirror 
to obtain the signal that is collected into the detector. 

In computing the surface integral, we define $\bm{r}$ as the position of the focus of the parabola, and $\bm{r}'$ as the generic point on the surface of the mirror. The current density depends on $\bm{r}'$ only through the retarded time $t_r$. As $|\bm{r}|\gg |\bm{r}'|$ and, as the integral is only computed on the surface of the mirror, we can rewrite
\begin{equation}
    \bm{E}(\bm{r},t)\approx \frac{1}{{r}}\int d\bm{s} \frac{\partial \bm{J}}{\partial t}(t_r).
\end{equation}
We use polar coordinates $(\rho,\theta)$ to take advantage of the geometry of the parabolic mirror. Being $F$ the focal distance of the parabola, we can rewrite the term $|\bm{r}-\bm{r}'|$ in the expression of $t_r$ as $|\bm{r}-\bm{r}'|=F+\rho^2/4F$, resulting in
\begin{equation}
    \bm{E}(\bm{r},t)\approx\frac{1}{r}\int_{0}^{2\pi}d\theta\int_0^{\bar\rho}  d\rho \,  \rho \frac{\partial\bm{J}}{\partial t}\Big( t-\frac{1}{c}\left(F+{\rho^2}/{4F}\right)\Big).
\end{equation}
$\bm{J}$ does not depend on $\theta$, so we can integrate out the angular variable, $\bar{\rho}$ is the radius of the mirror. Moreover, $\bm{J}$ is a smooth function of $t$, so the derivative can be taken out of the integral, and we change variable, defining $\xi=t-\frac{1}{c}(F+{\rho^2}/{4F})$ so that 
\begin{eqnarray}
   \bm{E} (\bm{r},t)  &\approx& \frac{2\pi}{r}\frac{\partial }{\partial t}\int_{\xi_{min}}^{\xi_{max}} d\xi \, 2cF \bm{J}(\xi) \nonumber \\
   &=& \frac{4\pi c F}{r}\frac{\partial}{\partial t}\left[\vec{\mathcal{J}}(\xi_{max})-\vec{\mathcal{J}}(\xi_{min})\right], 
\end{eqnarray}
with $\xi_{max}=\xi(\bar \rho)$ and $\xi_{min}=\xi(0)$
and $\partial{\vec{\mathcal{J}}}/\partial t =\bm{J}(t)$. Going back to the original variables we obtain
\begin{equation}
    \bm{E}(\bm{r},t)\approx \frac{4\pi cF}{r}\left\{\bm{J}\Big(t-\frac{1}{c}\Big(\frac{4F^2+\bar\rho^2}{4F}\Big)\Big)-\bm{J}\Big(t-\frac{F}{c}\Big) \right\}, 
\end{equation}
which we can rewrite as
\begin{equation}
    \bm{E}(\bm{r},t)\approx -\frac{4\pi cF}{r}\left[\bm{J}\left(t\right)-\bm{J}\left(t-T\right) \right].
\end{equation}

{Due to the integrating effect of the detector mirror the THz signal becomes proportional to the current $\bm{J}(t)$}{, i.e., similar to the initial pulse emitted at near field \eqref{seq:nearF}. However, we also notice a change in sign of the signal with respect to the initial pulse; this is consistent with the previously reported Gouy phase shift for THz emission \cite{wang2013, Ruffin1999PRL}.}
A secondary, back traveling, and opposite in sign, repetition of the signal appears with a time delay $ T=\bar\rho^2/4F$. If we consider a mirror of 100~mm focal length and 50~mm diameter, $T\approx5$~ps. 
{In general, in the case that $F\ll\bar\rho$ (i.e., short distances and/or large emitting surfaces) the time $T$ is so large that it effectively becomes infinite, removing the back-traveling peak from the measurements.
In the opposite limiting case, when we have $F\gg\bar\rho$ (i.e., large distances or small emitting surfaces) the time $T$ becomes short so that effectively the difference
$\bm{J}(t)- \bm{J}(t-T)$ becomes proportional to the time derivative, thus obtaining again the original $\partial \bm{J}/\partial t$ from Jefimenko's equation.}

\subsection{Off-axis mirror}
\label{app:off_mirr}
\begin{figure*}[htbp]
    \centering
\includegraphics[width=\linewidth]{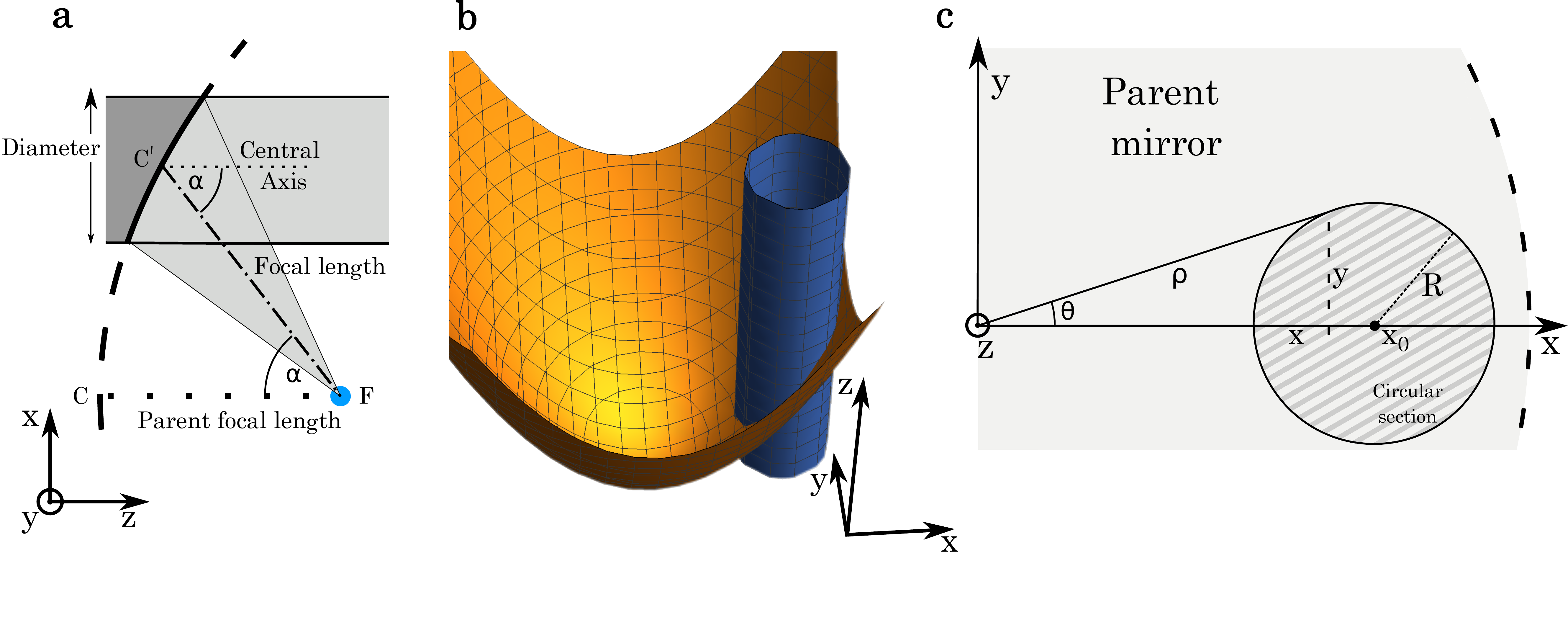}
\vspace*{-0.5cm}
    \caption{{Off-axis mirror as a section of a parabolic mirror.} (a) Schematics of an off-axis mirror in the $x-z$ plane, $\alpha$ is the off-axis angle. (b) 3D view of the circular section of a parabolic mirror. (c) Schematics of the circular section in the $x-y$ plane.}
\label{sfig:off-axis mirror}
\end{figure*}

The case of a centered mirror described in the previous section was analytically straightforward, but {adopted a somewhat} simplifying approximation. In such a setup the detector would screen the mirror from the incident radiation (that we suppose directed parallel to the mirror symmetry axis), and hence off-axis mirrors need to be used.   

As Fig.~\ref{sfig:off-axis mirror} shows, an off-axis mirror consists of a circular section cut out a parabolic mirror (parent mirror) that focuses the reflected radiation towards the focus of the original parabola from which the mirror is cut.
Relevant quantities in an off-axis parabolic mirror are:
\begin{itemize}
    \item the radius $R$ of the circumference that defines the circular section;
    \item the center of the mirror, i.e.\ the center of the circular section. We choose a coordinate system so that the center is in the point $(x_0,0,z_0)$;
    \item the effective focal length of the mirror, which is the distance between the center of the off-axis mirror and the focus of the original parabola;
    \item the focal length $F$ of the parent mirror;
    \item the angle between the optical axis of the parent mirror {and} the line connecting the focal point of the parent mirror with the center of the off-axis mirror. 
\end{itemize}
As per Fig.~\ref{sfig:off-axis mirror}(c), we consider here a 45° off-axis mirror, with a radius $R$ of 25~mm and focal length of 100~mm. We choose a coordinate system where the parent parabolic mirror of parent focal length $F$ is centered in the origin of the system, $F$ can be computed knowing the coordinates of the off-axis mirror and its focal length. The center of the off-axis mirror is located at position $x_0$ on the $x$ axis of the system. 

\begin{figure}[htbp]
    \centering
    \includegraphics[width=.85\linewidth]{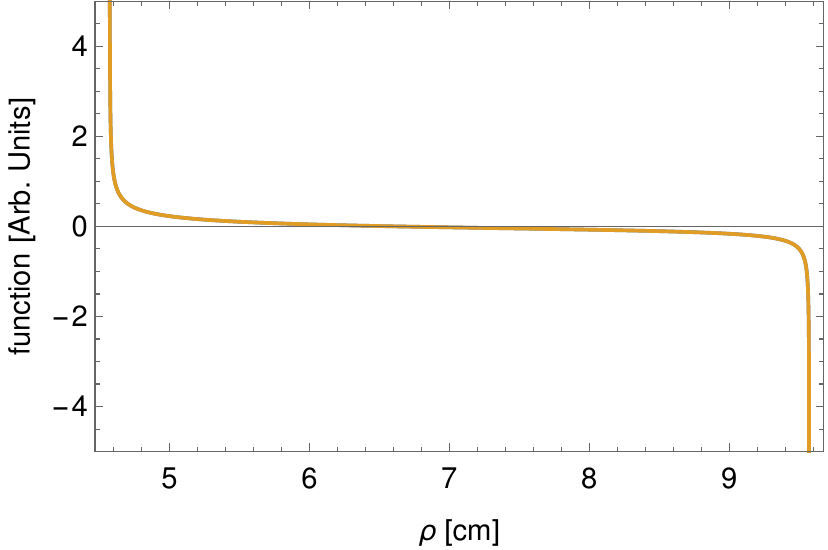}
    \caption{{Multiplying factor in the integral.} The factor with which  $\bm{J}$ is multiplied in Eq.~(\ref{seq:off_E}) plotted in the integrating interval, $x_0\pm x_R$, with $x_0\approx 7.1~\mathrm{cm}$. The function is divergent at the integral boundaries and almost flat and vanishing in the interval.}
    \label{sfig:peaks}
\end{figure}

The situation is less symmetric than in the previous case, the integral over the surface of the mirror has a different parametrization
\begin{eqnarray}
    \bm{E}(\bm{r},t) &\approx &\frac{1}{r}\int_{x_0-R}^{x_0+R}dx
\int_{-\sqrt{R^2-(x_0-x)^2}}^{\sqrt{R^2-(x_0-x)^2}}dy \nonumber \\
    && ~~\frac{\partial \bm{J}}{\partial t}\left[t-\frac{1}{c}\left(F+\frac{x^2+y^2}{4F}\right)\right] .
\end{eqnarray}
We can use polar coordinates as in the previous case, giving
\begin{equation}
    \bm{E}(t)\approx
    \int_{\rho_1}^{\rho_2}\rho \, d\rho
    \int_{\theta_1}^{\theta_2}d\theta
    \frac{\partial \bm{J}}{\partial t}\left[t-\frac{1}{c}\left(F+\frac{\rho^2}{4F}\right)\right] ,
\end{equation}
where, for the sake of simplicity we dropped the dependence on $\bm{r}$ in $\bm{E}$. 
The integration limits for the surface integral are
    $\rho_{1/2}=x_0\mp R$ and
\begin{equation}
\theta_{1/2}=\mp\arctan\left(\sqrt{\frac{4x_0^2\rho^2}{(\rho^2+x_0^2-R^2)^2}-1}\right), 
\end{equation}
and, after we integrate over $\theta$,
\begin{eqnarray}
    \bm{E}(t) \!&\approx&\!
    \int_{x_0-R}^{x_0+R}\rho \, d\rho\;
    2\arctan{\left(\sqrt{\frac{4x_0^2\rho^2}{(\rho^2+x_0^2-R^2)^2}-1}\;\right)}   \times
    \nonumber \\
    &&\frac{\partial \bm{J}}{\partial t}\left[t-\frac{1}{c}\left(F+\frac{\rho^2}{4F}\right)\right].
    \label{seq:rho_integ}
\end{eqnarray}
Next, we integrate by parts over $\rho$ and take into account that
\begin{equation}
    \frac{\partial \bm{J}}{\partial t}=\frac{\partial \bm{J}}{\partial \rho}\frac{\partial \rho}{\partial t}\qquad\mathrm{and}\qquad\frac{\partial \rho}{\partial t}=-\frac{1}{\rho}2Fc \, .
\end{equation}
\begin{widetext}
We can then rewrite Eq.~\eqref{seq:rho_integ} as
\begin{equation}
    -4Fc\int_{x_0-R}^{x_0+R}d\rho \, \arctan{\left(...\vphantom{\frac{A}{B}}\right)}
    \frac{\partial \bm{J} [...]}{\partial \rho}=\left.-4Fc\;\arctan{\left(...\vphantom{\frac{A}{B}}\right)}
    \bm{J}[...]\right|_{x_0-R}^{x_0+R}+4Fc\int_{x_0-R}^{x_0+R} d\rho \frac{\partial}{\partial \rho}\arctan{\left(...\vphantom{\frac{A}{B}}\right)}
    \bm{J}[...] ,
\end{equation}
where we omitted the arguments of $\arctan$ and of $\bm{J}$ for the sake of brevity. The first term in the right hand side is zero because the argument of the $\arctan$ is null at the integral boundaries. On the other hand, the derivative in the second term can be written explicitly as 
\begin{equation}
    \frac{\partial}{\partial \rho}\arctan{\left(\sqrt{\frac{4x_0^2\rho^2}{(\rho^2+x_0^2-R^2)^2}-1}\;\right)}=-\frac{R^2+\rho^2+x_0^2}{\rho\sqrt{
    4x_0^2\rho^2-(\rho^2+x_0^2-R^2)^2}\;}
\end{equation}
so that
\begin{equation}
   \bm{E}(t)\approx
   -4Fc\int_{x_0-R}^{x_0+R} d\rho\; \bm{J}\left[t-\frac{1}{c}\left(F+\frac{\rho^2}{4F}\right)\right] 
   \frac{R^2+\rho^2+x_0^2}{\rho \sqrt{
    4x_0^2\rho^2-(\rho^2+x_0^2-R^2)^2}\;}.
    \label{seq:off_E}
\end{equation}
In this integral, $\bm{J}$ is multiplied with a function that is strongly peaked around the integral boundaries and low-valued everywhere else, as shown in Fig.~\ref{sfig:peaks}. This means that the we can approximate the integral as $\bm{J}$ evaluated only at the peaks.
\end{widetext}

Finally, the result of the integral can be written as a linear combination of $\bm{J}$ evaluated at the peaks (the boundaries of the integral), giving 
\begin{eqnarray}
    \bm{E}(t)&\approx &a \, \bm{J}\left[t-\frac{1}{c}\left(F+\frac{(x_0+R)^2}{4F}\right)\right]+ \nonumber \\
    && b \, \bm{J}\left[t-\frac{1}{c}\left(F+\frac{(x_0-R)^2}{4F}\right)\right] ,
\end{eqnarray}
with \begin{equation}
    \frac{a}{b}=-\frac{4x_0(x_0+R)-1}{4x_0(x_0-R)-1}
\end{equation}
which is negative and greater than 1 in modulus, meaning that the two peaks are opposite in sign, as in the previous case with the centered mirror. However, here the two peaks have different values, which is consistent with the less symmetric geometry of the mirror. 

The time separation between the two peaks is $T=x_0R/(Fc)$, which is approximately $\approx$20~ps in the 15$^\circ$ case and $\approx$70~ps in the 45$^\circ$ case, compared with the centered mirror case the separation of the two peaks is much larger, again due to the lower symmetry of this system. Furthermore, as we stated for the centered mirror case, 
after integrating over the surface of the mirror, the electric field becomes proportional to the charge current density $\bm{J}$. 

As a last remark, we point out that the factor multiplying $\bm{J}$ in Eq.~\eqref{seq:off_E} in the integral has, contrary to e.g.\ a Dirac $\delta$-function, peaks of finite width meaning that after the integration the profile of $\bm{J}$ will be slightly distorted, as can be seen  in Fig.~\ref{fig:off-axis_signal}.   

\section{Response function of the detector crystal}
\label{app:response}

In this section {we summarize  the  theory of THz radiation detection through electro-optic crystals \cite{gallot1999electro,casalbuoni_numerical_2008,wu_comparison_2014}. {Specifically,} 
we compute the response function of the detecting crystals used int THz experiments}. 

The detection of a THz pulse in a crystal is performed by propagating collinearly both the THz pulse and a probe pulse in the crystal. We consider here the case where the 800~nm pulse that excites the THz emission is split to be used as a probe pulse as well. As the THz pulse propagates in the crystal, it induces birefringence in it, which affects the propagation of the optical probe pulse. When both pulses propagate in the crystal, the polarization of the optical pulse rotates. By tracking the rotation of the polarization of the probe pulse the THz signal is reconstructed. 

To model the response of the crystal to the presence of the THz and probe pulses, we need to know the crystal optical properties, i.e., we need information about the crystal refractive index and extinction coefficient.
We model the dielectric constant of the crystal with the Lorentz oscillator model 
\begin{equation}
\epsilon(\omega)=\epsilon_{el}+\frac{S_0\omega_0^2}{\omega_0^2-\omega^2-i\Lambda_0\omega}.
\end{equation}
Here, the first term $\epsilon_{el}$ represents the contribution to the dielectric constant of bound electrons; it is constant in the THz range. The second term is the contribution from the lowest transverse-optical lattice oscillation in the crystal, treated as a damped harmonic oscillator (contributions from higher order modes have been neglected). ${S}_0$ is the strength of the oscillator, $\omega_0$ its eigenfrequency and $\Lambda_0$ the damping constant. The parameters for the ZnTe crystal that is commonly used in these experiments, are reported in Table~\ref{tab:Xparam}. The refractive index of the crystal $n$ and its extinction coefficient $\kappa$ are,  respectively, the real and imaginary part of $\epsilon^{1/2}$.

The refractive index of the crystal is frequency dependent. Due to this, the THz and probe pulse propagate through the crystal with different velocities. Therefore, the two pulses experience a phase mismatch (for convenience, the following quantities are expressed in term of the frequency $f=\omega/2\pi$)
\begin{equation}
    \Delta k(f)=\frac{2\pi f}{c}\left(n(f)-n_g(f_\mathrm{probe})\right),
\end{equation}
where $n_g$ represents the refracting index experienced by the probe signal in the crystal. The optical probe pulse travels with group velocity
 \begin{equation}
    v_g(f)=\frac{c}{n(f)+fn'(f)},
\end{equation}
from which we can define the group refractive index as $n_g=c/v_g$, with $c$ being the speed of light. 

\begin{figure}[htbp]
    \centering
    \includegraphics[width=.85\linewidth]{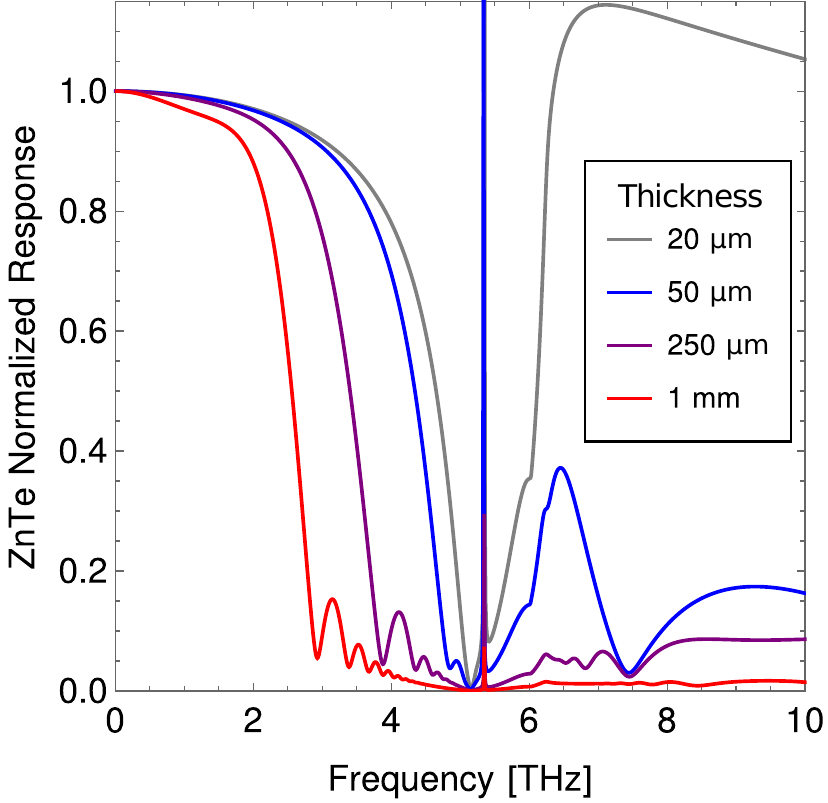}
    \caption{{ZnTe response function.}
    The response function of the ZnTe crystal has been computed for different thicknesses of the detector crystal and for a wavelength of the probe pulse of 800~nm.}
    \label{sfig:ZnTe}
\end{figure}

It follows that, as the two pulses propagate through a crystal with thickness $d$, the mismatch between the two increases, limiting the frequency window in which the probe pulse and the THz pulse can interact. By integrating the phase mismatch over the thickness of the crystal, we obtain the \textit{geometric} response function of the crystal
\begin{equation}
    G(f)=\frac{1}{d}\int_0^d dz \exp{\left( i\Delta k(f)z-\frac{2 \pi f}{c}\kappa z\right)}.
\end{equation}
However, to obtain the electro-optical response function of the crystal, we must take into account also the frequency-dependent transmission properties of the crystal, summarized by the Fresnel transmission coefficient 
\begin{equation}
    T(f)=\frac{2}{1+n(f)+i\kappa(f)}. 
\end{equation}
Finally, merging the geometric response function and the transmission coefficient, we can define the electro-optical response function as 
\begin{equation}
    S_{EO}(f)\approx r_{41}(f)
    G(f)T(f), 
    \label{seq:SEO}
\end{equation}
where $r_{41}$ is the electro-optical coefficient. This coefficient represents the shift in the refractive index of a medium when an electric field is applied to it. It can be modeled similarly with Lorentz oscillators as the dielectric function,
\begin{equation}
    r_{41}(\omega)=d_E\left(1+\frac{C \omega_0^2}{\omega_0^2-\omega^2-i\Lambda_0 \omega}\right),
\end{equation}
where $r_{41}(f)=r_{41}(\omega/2\pi)$ and $d_E$ and $C$ are empirically-determined coefficients and are reported in Table~\ref{tab:Xparam}.

Figure~\ref{sfig:ZnTe} shows the response function for a ZnTe crystal computed with Eq.\ \eqref{seq:SEO} and the parameters from Table~\ref{tab:Xparam} for different values of the crystal thickness $d$, normalized for easier visualization to the value of the response at zero frequency. We observe that the bandwidth of the detector quickly decreases with the thickness of the crystal. Moreover, the response shows oscillations and a resonance peak around the value $\omega_0/2\pi$, the frequency of the lowest transverse-optical lattice oscillation. This resonance peak is still present in the real response function, {however it should be less pronounced as the contributions of higher order lattice oscillations mitigate the divergent behavior at $\omega_0$. }

Finally, we point out that the value of $n_g(f_{probe})=3.1$ that we used to compute the response function for ZnTe in Fig.~\ref{sfig:ZnTe}
has been increased by {about}
10\% with respect of the value obtained by using the reported parameters ($n_g^{ZnTe}=2.72$), {as this appeared more compatible with experimental findings.}

\begin{table}[h]
\begin{ruledtabular}
 \caption{Model parameters for the electro-optical response of ZnTe crystals. The values given are taken from Ref.\ \cite{casalbuoni_numerical_2008}.}
    \centering
    \begin{tabular}{c c c c c c c}
    \vphantom{$\int$}
       & $\epsilon_{el}$ & S$_0$ & $\omega_0/2\pi$  & $\Lambda_0/2\pi$ & $d_E$ & $C$ \\
    \vphantom{$\int_a^b$}
       & & & [THz] & [THz] & [m/V] & \\[0.2cm]\hline \\[-0.3cm]
    ZnTe & 7.4 & 2.7 & 5.35 & 0.09 & 4.25 10$^{-12}$& -0.07 \\
    \end{tabular}
   \end{ruledtabular}
    \label{tab:Xparam}
\end{table}

\begin{acknowledgments}
We thank Branislav Nicoli{\'c}, Tobias Kampfrath, Henri Jaffr{\`e}s and Hjalmar Lindstedt for helpful discussions. This work has been supported by the Swedish Research Council (VR), the German Research Foundation (Deutsche Forschungsgemeinschaft) through CRC/TRR 227 ``Ultrafast Spin Dynamics'' (project MF, project-ID: 328545488)), the K.\ and A.\ Wallenberg Foundation (Grants No.\ 2022.0079 and 2023.0336), and the European Union’s Horizon 2020 Research and Innovation Programme under FET-OPEN Grant Agreement No.\ 863155 (s-Nebula). The computational resources were provided by the National Academic Infrastructure for Supercomputing in Sweden (NAISS) at NSC 
Link\"oping, partially funded by VR through Grant Agreement No.\ 2022-06725.
\end{acknowledgments}

\bibliography{bibliography}

\end{document}